\documentclass[useAMS,usenatbib,usegraphicx]{mn2e}

\title{A transient relativistic radio jet from Cygnus X-1} 
\author[R. P. Fender et al.]
       {R. P. Fender$^1$, A.M. Stirling$^2$
R. E. Spencer$^2$,  I. Brown$^2$, G. G. Pooley$^3$, \newauthor
T. W. B. Muxlow$^2$, J.C.A. Miller-Jones$^4$\\
$^1$ School of Physics and Astronomy, University of Southampton, Southampton SO17 1BJ\\
$^2$ University of Manchester, Jodrell Bank Observatory, Macclesfield, Cheshire SK11 9DL\\ 
$^3$ University of Cambridge, Cavendish Laboratory, Madingley Road, Cambridge CB3 0HE\\
$^4$ Astronomical Institute `Anton Pannekoek', University of Amsterdam, Kruislaan 403, 1098 SJ Amsterdam, The Netherlands\\
}
\date{Accepted 0000.
      Received 0000
}
\pagerange{\pageref{firstpage}--\pageref{lastpage}}
\pubyear{2005}
\begin{document}
\label{firstpage}
\maketitle

\begin{abstract}
We report the first observation of a transient relativistic jet from
the canonical black hole candidate, Cygnus X-1, obtained with the
Multi-Element Radio-Linked Interferometer Network (MERLIN). The jet
was observed in only one of six epochs of MERLIN imaging of the source
during a phase of repeated X-ray spectral transitions in 2004
Jan--Feb, and this epoch corresponded to the softest 1.5-12 keV X-ray
spectrum.  With only a single epoch revealing the jet, we cannot
formally constrain its velocity. Nevertheless, several lines of
reasoning suggest that the jet was probably launched 0.5--4.0 days
before this brightening, corresponding to projected velocities of
$0.2c \la v_{\rm app} \la 1.6c$, and an intrinsic velocity of $\ga
0.3c$.  We also report the occurrence of a major radio flare from Cyg
X-1, reaching a flux density of $\sim 120$ mJy at 15 GHz, and yet not
associated with any resolvable radio emission, despite a concerted
effort with MERLIN. We discuss the resolved jet in terms of the
recently proposed 'unified model' for the disc-jet coupling in black
hole X-ray binaries, and tentatively identify the 'jet line' for Cyg
X-1. The source is consistent with the model in the sense that a
steady jet appears to persist initially when the X-ray spectrum starts
softening, and that once the spectral softening is complete the core
radio emission is suppressed and transient ejecta / shock observed.
However, there are some anomalies, and Cyg X-1 clearly does not behave
like a normal black hole transient in progressing to the canonical
soft / thermal state once the ejection event has happened.
\end{abstract}


\section{Introduction}

\begin{figure*}
\begin{center}
\includegraphics[width=12cm,angle=270]{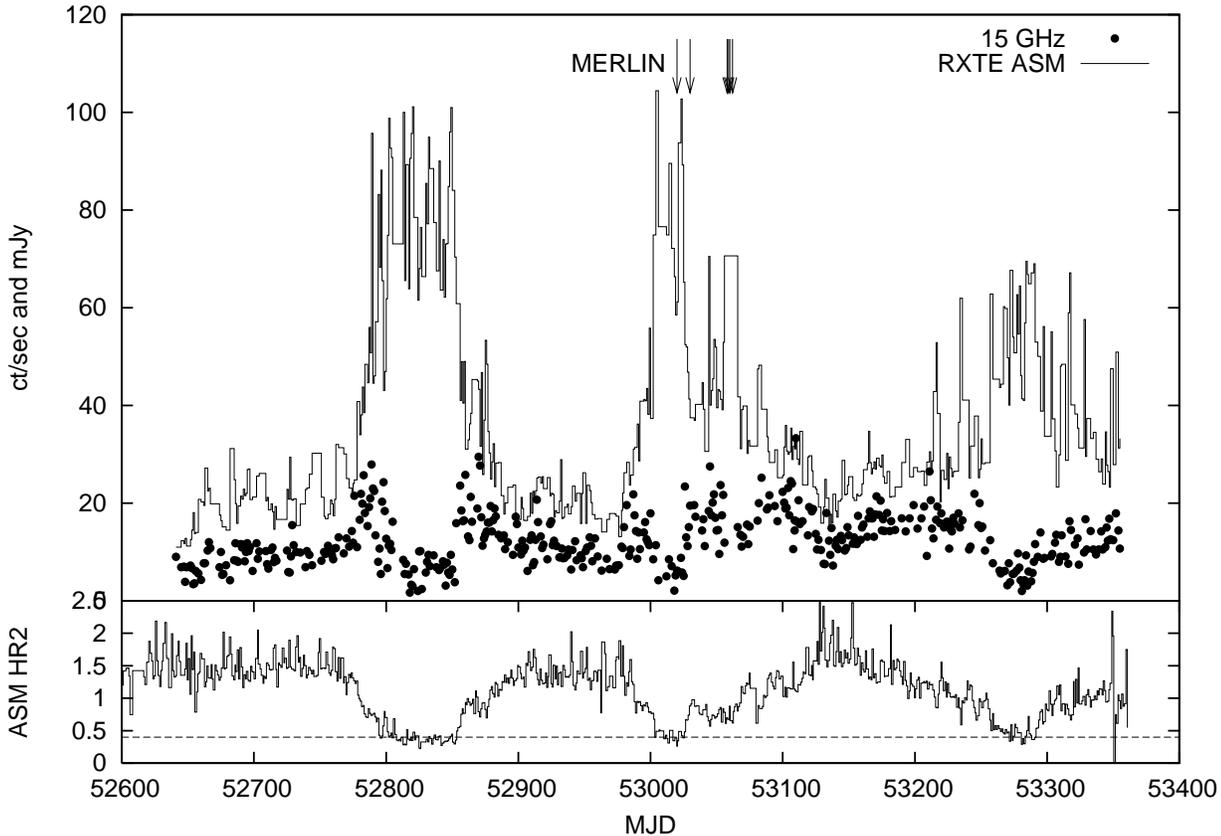}
\end{center}
\caption{Radio and X-ray monitoring of Cygnus X-1, from the Ryle
Telescope and RXTE ASM. All data points are daily averages. 
The top panel shows the radio and X-ray fluxes, and the lower
panel the ratio of counts in the (5--12) / (1.5--3) keV bands;
higher values indicate 'harder' X-ray spectra. Clear
periods of radio:X-ray correlations (hard X-ray state) and
anticorrelations (soft[er] X-ray state) are evident. The epochs of
MERLIN observations are indicated by arrows at the top of
the figure. Only the first MERLIN run, when the core radio source was
very faint, revealed a jet.  }
\end{figure*}

Relativistic jets (e.g. Hughes 1991; Burgarella, Livio \& O'Dea 1993;
Livio 1996) are a common feature of accretion onto compact objects,
most notably black holes, on all scales.  They are important channels
for the removal of gravitational potential and/or spin energy, as well
as angular momentum, and yet the details of their formation remain
largely unclear. Recent studies of stellar-mass ($3 M_{\odot} \la M_{\rm
BH} \la 15 M_{\odot}$) black holes in binary systems have led to some
empirical understanding of the coupling between accretion 'states' and
the jet formation process (Fender, Belloni \& Gallo 2004 and
references therein). Quantitative scalings between these low-mass
black holes and the supermassive black holes in distant galaxies
(e.g. Merloni, Heinz \& di Matteo 2004; Falcke, K\"ording \& Markoff
2005) have finally given firm grounding to the hope that by studying
X-ray binaries we may further our understanding of active galactic
nuclei.

Cygnus X-1 (HDE 226868, V1357 Cygni) comprises a supergiant secondary
(spectral type O9.7 Iab) with mass between 20--33 M$_{\odot}$,
together with a compact primary which is a strong black-hole candidate
with mass between 7--16 M$_{\odot}$ (e.g. Gies \& Bolton 1986a). The
system spends most of its time in a bright but hard X-ray state, with
a bolometric X-ray luminosity around 2\% of Eddington (e.g. di Salvo
et al. 2001).

High angular resolution Very Long Baseline Array (VLBA) + phased Very
Large Array (VLA) observations from 1998 Aug show a clearly resolved
and persistent jet over three closely spaced epochs in the low/hard
state (Stirling et al. 2001). This was the first direct imaging of
steady radio jet from a source in the hard X-ray spectral state. More
recently, Gallo et al. (2005) have discovered a large-scale ($\geq 5$
arcmin) radio and optical lobe approximately orientated with the
mas-scale jet. Assuming the lobe is the result of the prolonged action
of the jet on the ambient medium, Gallo et al. (2005) were able to
demonstrate that the jet must be carrying a time-averaged power
comparable to the present-day X-ray luminosity.

\section{Observations}

\subsection{Ryle Telescope and RXTE ASM monitoring}

The Ryle Telescope (RT) is used from time to time for monitoring
variable sources, including Cygnus X-1, at 15 GHz. The basic
parameters of the RT and its monitoring program are described in
Pooley \& Fender (1997). The approximately daily monitoring of Cygnus
X-1 with the RT is complemented by data from the Rossi X-ray Timing
Explorer (RXTE) all sky monitor (ASM). In addition to tracking the
evolution of the total X-ray flux of the source, the RXTE ASM data
also allow us to measure hardness ratios, indicative of the X-ray
'state' of the source (e.g. Fender, Belloni \& Gallo 2004; Remillard
\& McClintock 2005; Homan \& Belloni 2005; Belloni
2005). Daily-averaged data from the RT and RXTE ASM, centred on the
period of our MERLIN observations (see below) are presented in Fig 1;
see also Figs 3--6.

\subsection{MERLIN}

MERLIN was initially triggered to observe Cyg X-1 by a change in the
X-ray behaviour of Cyg X-1 to a generally brighter and more variable
state, commencing around 2003 December (see Fig 1). This trigger
resulted in observations at epochs M1 and M2 (see Table 1). A major
radio flare detected by the RT on 2004 Feb 20 resulted in a second
MERLIN trigger and observations at epochs M3-6.  We observed
while the array was undergoing engineering work and so the specific
telescopes observing changed between epochs. Altogether we obtained
varying amounts of data over 8 separate days. All of the MERLIN
observations were taken at 4.994 GHz and with 16 MHz bandwidth in each
of the LL, RR, LR and RL polarization products.

The epochs of the MERLIN observations are marked on Fig 1. From 2004
Feb 20 -- 2004 Feb 22 Cambridge was involved sporadically (about 50,
10 and 30 per cent respectively of each days' observing time) and
Knockin had a warm receiver cryostat throughout, it was removed from the
array at 2004 Feb 23 10:00. While available Knockin was included,
although noisy, to improve the {\it{(u, v)}} plane coverage.

The data reduction was done in the standard way for continuum mapping,
outlined in the MERLIN user guide {\footnote{available at {\bf
http://www.merlin.ac.uk/}}}.  We used 3C 286 as a flux density and
polarisation position-angle calibrator, OQ 208 or 0552+398 as a point
source calibrator. The approximately 400 mJy source 1951+355, which is
unresolved to MERLIN at this frequency, was used as a phase reference
and to calibrate the d-terms. The useful bandwidth was 14 MHz after
editing the end channels.

Amplitude and phase self-calibration was performed on the phase
reference source (shifting it to the phase-centre) and interpolated to
the target, Cygnus X-1. Observations taken on Cygnus X-1 and 1951+355
were flagged at elevations beneath 15$^{\circ}$ as phase and amplitude
calibration transfer is generally poorer at low elevations. The
imaging was performed using {\tt {IMAGR}} within AIPS (Diamond
1995). A preliminary image showed that our coordinates used in the
correlation for either the phase-reference or target source were
slightly in error ($\sim$30 mas north-east of the actual source
position for Cygnus X-1) and the phase centre was shifted with {\tt
{UVFIX}} within AIPS to the target position. As the proper motion of
Cygnus X-1 on timescales of months is very small compared to our
resolution this shift was applied to all of the epochs. From these
observations our best coordinates for Cygnus X-1 are (J2000) RA 19 58
21.6747 Dec 35 12 05.767; the peak core flux densities listed in table
1 were all measured within 4 mas of this position.

\begin{figure}
\includegraphics[width=8.5cm]{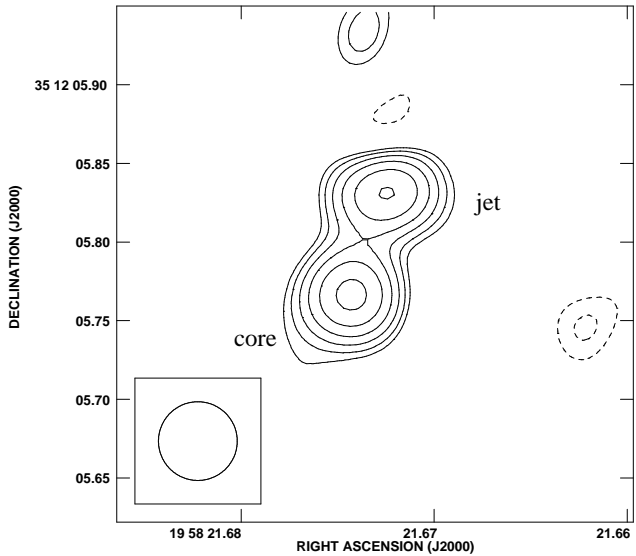}
\caption{A resolved transient jet from Cygnus X-1. This image
corresponds to epoch M1 when the core was very faint, and is the only
one of six MERLIN epochs reported here in which the source was
resolved. The contour levels are indicated on the figure.  Contours
are at levels of (-2, -1.41, -1, 1, 1.41, 2, 2.828, 4, 5.657, 8,
11.31, 16) $\times 0.2$ mJy / beam. The core and jet peak fluxes in
this image are 1.24 ± 0.06 and 1.18 ± 0.06 mJy / beam respectively;
the values listed in table 1 are for the entire of epoch M1.}
\label{mer2}
\end{figure}

%The problems involved in synthesis imaging of intrinsically varying
%sources are well known. In an attempt to avoid these we selected
%regions in time with approximately (within 20 percent of the mean)
%constant flux densities. 

All of the images were restored with a 50 mas circular beam. Table 1
summarizes the results from each image. All epochs apart from epoch
M1, were found to be point sources (limits on the flux from jets are
given in Table 1). No linear polarization was detected from the source
at any time; the most stringent limit being $P / I < 2.3$\% $(3
\sigma)$.

In Fig 2 we present the image from epoch M1, the only epoch at which a
resolved jet was detected, and also the epoch with by far the weakest
core emission.  The source is resolved towards the north-west with an
apparent bright component, possibly superposed on a smooth underlying
jet. Applying the AIPS task {\tt {JMFIT}} we find two approximately
beam-sized Gaussian components of integrated intensities 1.6 mJy
(core) and 0.7 mJy (jet); see table 1. The position angle between the
centres of these Gaussians is -25$^{\circ}$ $\pm$ 3$^{\circ}$. The
angular separation between core and jet is $70 \pm 5$ mas
(corresponding to $0.82 \pm 0.06$ light days at a distance of 2 kpc).

For comparison, epoch M6 is our image with the highest dynamic range
after the 120 mJy flare of 2004 Feb 20. We have used an identical beam
size and contouring scheme as used for epoch M1. There is no trace of
extension in any position angle. To stress the point,
had the extended 'jet' observed at epoch M1 been present in any of the
other epochs, it would have been detected -- its detection at epoch M1
is not just a consequence of the weaker core at this time.

In an attempt to pin down the time of appearance of the extended radio
emission, we have broken epoch M1, which was a $\sim23$\,hr run from
start to finish, into different segments.  A significant gap in the
observations, between 2004 Jan 15 21:00 - Jan 16 03:00 separates the
data into two segments, albeit of different length (6 and 15 hr
respectively).  The jet was only detected in the second of the two
blocks of data.  However, since MERLIN does not have enough baselines
to operate as a `snapshot' type of instrument (unlike the VLA), the
\textit{uv}-coverage of the two blocks would have been very different.
We therefore tested whether the jet appearing in the second half of
the data would have been detected by the {\it uv}-coverage of the
first half.  The clean components from the jet in the image of the
second half were selected using the AIPS task {\sc ccedt} and added
back in to the {\it uv}-data of the first half using the task {\sc
uvsub}.  Imaging this modified {\it uv}-data set, it was not possible
to unambiguously detect the jet above the noise. A parallel approach
was used in MIRIAD (Sault, Teuben \& Wright 1995) with the task
UVMODEL, and the same conclusions reached.  From these data, we cannot
therefore place a limit on the jet velocity based on epoch M1 alone.

\begin{table}
\begin{center}
\begin{tabular}{|c|c|c|c|}
\hline
Epoch           & Date (2004) & Core peak& Jet peak\\
       &  & (mJy bm$^{-1}$) & (mJy bm$^{-1}$) \\
\hline
M1 & Jan 15/16      & $1.2 \pm 0.07$  &  $1.2 \pm 0.07$         \\
M2 & Jan 26              & $11.8 \pm 0.17$ &  $<0.51$         \\
M3 & Feb 21              & $9.9 \pm 0.20$  &  $<0.60$        \\
M4 & Feb 22              & $6.9 \pm 0.28$  &  $<0.84$        \\
M5 & Feb 23              & $7.8 \pm 0.30$  &  $<0.90$        \\
M6 & Feb 25              & $13.6 \pm 0.13$ &  $<0.39$         \\
\hline
\end{tabular}
\end{center}
\caption[Summary of MERLIN image parameters] 
{Summary of the MERLIN observations of
Cygnus X-1. }
\label{mertab}
\end{table}

\section{Discussion}

\begin{figure}
\begin{center}
\includegraphics[width=6cm,angle=270]{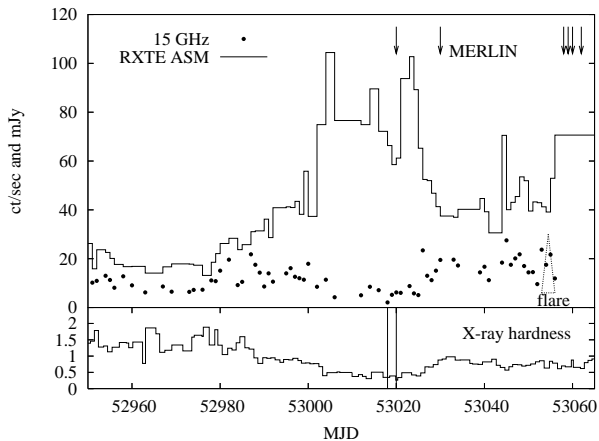}
\end{center}
\caption{Radio and X-ray monitoring around the time of the MERLIN
observations.  The top panel presents the Ryle Telescope 15 GHz flux
densities and the total (2-12 keV) X-ray count rate from the RXTE
ASM. Arrows indicate the times of the MERLIN observations; only the
first observation (epoch M1, see text) detected an extended jet. The
dotted triangle in the upper panel indicates the time of the large
radio flare plotted in Fig 4, which corresponded to a large increase
in the X-ray flux. The lower panel indicates the hardness ratio from
the ASM (5--12 keV/1.5--3 keV), and the vertical lines indicate our
best estimate of the range of launch times for the jet which
brightened during epoch M1.}
\end{figure}

\begin{figure}
\begin{center}
\includegraphics[width=5.5cm,angle=270]{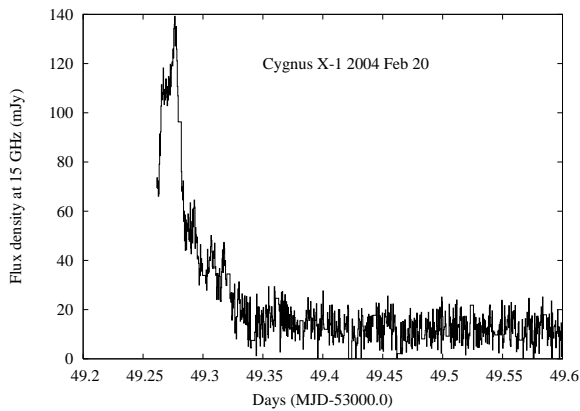}
\end{center}
\caption{A close up of the radio flare observed at 15 GHz with the RT
on 2005 Feb 20, with 32-sec sampling. This is the highest radio flux
density recorded from the source with the RT to date. MERLIN epochs
M3-6 were performed between 1--5 days after this event, and yet none
of them resolved a discrete ejection event, unlike epoch M1.}
\end{figure}

\subsection{Disc-jet coupling}

Fender, Belloni \& Gallo (2004) have outlined a model in which
discrete radio events are associated with transitions from `hard'
X-ray spectral states to `softer' states at relatively high ($\geq
1$\% Eddington) luminosities. In the accretion flow this seems to be
associated with a rapid inwards transition by the optically thick,
geometrically thin, accretion disc. In the outflow it appears to
correspond to a transition between a state which produces a steady jet
and one which produces no jet, via a brief phase in which the jet
Lorentz factor increases, resulting in internal shocks in the flow. It
may be these internal shocks which we identify as discrete components
in radio maps of X-ray transients (see also Kaiser et al. 2000).

The behaviour of Cyg X-1 appears to follow a consistent pattern,
although with some anomalies. Figure 3 presents a close-up of the data
around the time of the one resolved event, 2004 Jan 15/16. Precisely
around the time of the MERLIN observation the radio flux dropped to
very low levels, towards the end of a period of X-ray spectral
softening.  The radio flux stayed at these very low ($< 2$ mJy at 5
GHz and undetectable at 15 GHz) levels for 12--18 hr and then
recovered rapidly recovered to previous levels 6--12 hr before the
X-ray spectrum returned to a hard state. Unfortunately the RXTE ASM
sampling was rather sparse around this time and this hampers our
precise interpretation of the sequence of events.

\begin{figure}
\begin{center}
\includegraphics[width=6.5cm,angle=270]{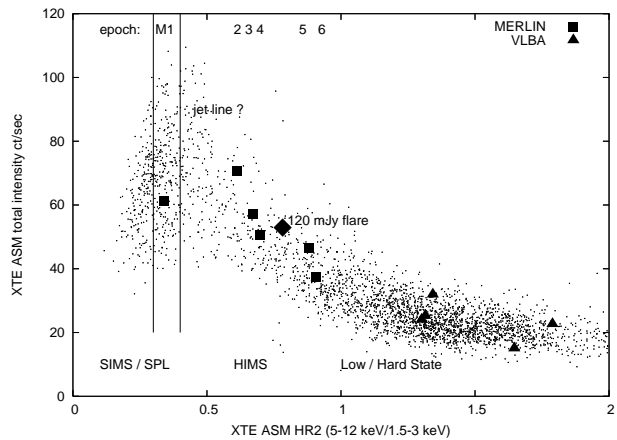}
\end{center}
\caption{A hardness intensity diagram for Cyg X-1, based upon the RXTE
ASM one-day averages.  The epochs of the MERLIN observations, previous
VLBA observations which revealed a compact ($\leq 20$ mas) steady jet,
and the large radio flare observed by the RT (Fig 4), are
indicated. The observation of the jet is the only one to the left of
the peak which we might tentatively associate with the `jet line' in
this source. However, this interpretation does not explain the nature
of the 2004 Feb 20 flare (Fig 4), which appears to have occurred to
the right of the line, yet we would naturally interpret as being
associated with an ejection / shock event. 'HIMS' = 'Hard Intermediate
State'; 'SIMS' = 'Soft Intermediate State'; 'SPL' = 'Steep Power Law'
-- see Fender et al. (2004); Homan \& Belloni (2005); McClintock \&
Remillard (2005); Belloni (2005).}
\end{figure}

\begin{figure}
\begin{center}
\includegraphics[width=5.5cm,angle=270]{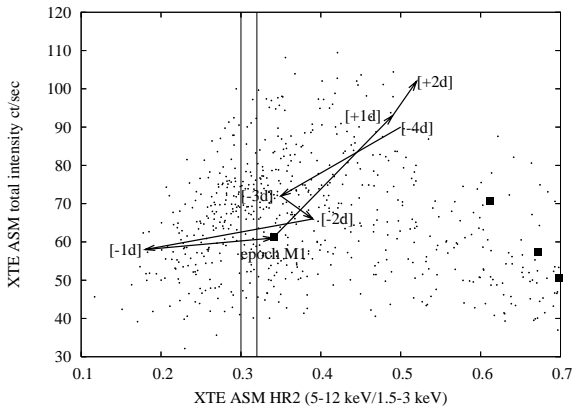}
\end{center}
\caption{Motion of Cyg X-1 in the HID for four days preceding and two
days following the jet which was imaged at epoch M1.}
\end{figure}

Fig 3 also shows that the radio flare of 2004 Feb 20 (Fig 4) occurred
around the time of a large step-like increase of the integrated X-ray
flux, while the source was in a `hard intermediate' state (Fender et
al. 2004; Homan \& Belloni 2005). Fig 5 (see also Fig 6) sheds more
light on this in the form of a hardness-intensity diagram (HID) for
Cyg X-1. There is a clear peak in the luminosity distribution of the
source, corresponding to a hardness ratio HR2 (counts in 5--12 keV
band / counts in 1.5--3 keV band) around 0.3--0.4. In the canonical hard
state $1.0 \la$HR2$\la 1.5$ is more typical. It is interesting to note
that epoch M1 -- the only epoch at which a jet was imaged -- is the
only epoch to the left of (softer than) the peak. All of the other
MERLIN epochs correspond to `hard intermediate' states (HIMS; Belloni
2005) with $0.5 \la$HR2$\la 1.0$. We therefore tentatively associate
HR2$\sim 0.4$ with the `jet line' (Fender et al. 2004) for Cyg
X-1. This is consistent with the picture of Fender et al. (2004) and
Corbel et al. (2004) in which the steady jet (core emission) persists
during softening from the canonical hard state, until some critical
point (represented spectrally by the `jet line') is reached. Also
marked in Fig 5 are the state of Cyg X-1 at epochs when a steady,
compact ($\leq 20$ mas) jet was imaged with the VLBA (Stirling et
al. 2001; R. Spencer private communication); these are all in the
canonical low/hard state.

Fig 6 shows the evolution of daily-averaged counts for Cyg X-1 in the
HID for the four days preceding and two days following the imaging of
the jet on 2004 Jan 15/16. Based on this more detailed analysis, we
find that the jet line is likely to lie close to HR$\sim$0.3; this is
indicated on the figure.  We note that Cyg X-1 exhibits X-ray temporal
and spectral variability on much shorter timescales than those
utilised here from the RXTE ASM, and we may be missing key events by
using daily averages. It is also clear that Cyg X-1 does not behave
like a `typical' transient once it has crossed the jet line, in
progressing towards the canonical soft / thermal state, but instead
returns to the hard IS state within two days. The only other source to
exhibit such behaviour is GRS 1915+105, although it is likely that
repeatedly radio-bright objects such as Cygnus X-3 may behave
similarly.

It is also interesting to speculate about the nature of the core
emission at epoch M1.  It has been well established for some years now
(e.g. Fender et al. 1999) that the radio emission from black hole
X-ray binaries is suppressed in soft X-ray states compared to hard
states. An unambiguous detection of core (i.e. currently generated
flow) radio emission from a source in a steady soft state has not yet
been reported. If the `core' radio emission we detect at epoch M1 is
indeed the supressed level, then the quenching in Cyg X-1 is only a
factor of $\sim 10$, compared to $\ga 30$ and $\ga 50$ in the black
hole X-ray binaries GX 339-4 and XTE J1550-564 respectively (Fender et
al. 1999; Corbel et al. 2001). However, there is no way of knowing if
this weak core is completely or only partially quenched given, again,
the known rapid variability of Cyg X-1.  What we observe as the core
could in fact be emission from a receding jet which in many cases will
have a very small proper motion (much less than that of the
approaching jet). In this scenario the jet emission would have to have
some flux evolution including a local peak; for a symmetric
monotonically decaying event the receding jet cannot appear brighter
than the approaching.

\subsection{Jet launch and velocity}

As discussed above, it is very difficult to pin down the moment at
which point the matter / energy associated with the resolved jet was
launched. In the model of Fender, Belloni \& Gallo (2004) the launch
moment would correspond to the end of a phase of X-ray spectral
softening and would be immediately followed by a quenching /
suppression of the core radio emission. However, given the limited
quality of data from the RXTE ASM, even for a very bright source such
as Cyg X-1, we can only say that $0 \la \Delta t \la 15$ where $\Delta
t$ was the time since launch in days. The resulting associated proper
motion and velocity would be $\mu = 70 / \Delta t$ mas d$^{-1}$ and
$v_{\rm app} = 0.8c / \Delta t$. The more detailed inspection of the
HIDs outlined above suggests - no more - that the jet was launched at
most $\sim 4$ days before epoch M1. This interpretation is supported
circumstantially by the very rapid decay of the large flare event on
2005 Feb 20 (Fig 4). Furthermore, we estimate that proper motions
larger than $\sim 100$ mas d$^{-1}$ would probably have caused
smearing in the M1 image which was not observed.

For $0.5 \la \Delta t \la 4$ days, $140 \la \mu \la 17.5$ mas
$d^{-1}$, $1.6c \la v_{\rm app} \la 0.2c$. The Cyg X-1 system has
been extensively modelled based on optical and UV data, from which a
general consensus has emerged that the orbital inclination angle of
the system is $\sim 30^{\circ}$ (e.g. Gies \& Bolton 1986b). If the
jet shares the same inclination angle (and this is by no means certain
-- Maccarone 2002) then the true velocity of the jet would be $\geq
0.3$c (given the accumulation of uncertainties, we cannot place an
upper limit on the Lorentz factor of the jet -- see Fender 2003;
Miller-Jones, Fender \& Nakar 2005 for further discussion).

\section{Conclusions}

We report the first observation of a transient, extended (70 mas)
radio jet from the classical black hole binary Cygnus X-1. The jet was
observed in only one of six MERLIN observations performed while the
system was undergoing frequent X-ray state changes. The one epoch in
which the jet was imaged was that with the softest 1.5--12 keV
spectrum; at that moment the core radio emission was detectable but a
factor $\sim 10$ weaker than normal. Estimates of the launch epoch of
the jet indicate it probably has a bulk velocity $v \ga 0.3c$, with
essentially no formal upper limit.

At the other five epochs the source X-ray spectrum was harder,
although still not in the canonical hard state, and displayed more
powerful core emission and no extended jet. Three of these
observations were made within days of the brightest radio flare
observed from Cyg X-1 ($\sim 120$ mJy at 15 GHz), and yet no extended
radio emission was detected. It is not clear if this may be related to
the jet expanding into a rarefied bubble (Gallo et al. 2005), or if
for some reason the flare was, unusually, not associated with a
jet. Throughout the interpretation of the X-ray:radio behaviour, we
have to bear in mind that Cyg X-1 varies on much shorter timescales
than the RXTE ASM sampling, and we may have missed some key event.
Nevertheless, the overall pattern of behaviour seems to be broadly
consistent with the `unified' model for black hole X-ray binary jets
put forward by Fender, Belloni \& Gallo (2004) although there are some
clear anomalies.

\section*{Acknowledgments}

RPF would like to thank Tom Maccarone for comments on an early draft
of the manuscript, and the referee for detailed and constructive
criticism.  MERLIN is operated as a National Facility by the
University of Manchester at the Nuffield Radio Astronomy Laboratories,
Jodrell Bank, on behalf of the Particle Physics and Astronomy Research
Council (PPARC). We thank the staff at MRAO for maintenance and
operation of the Ryle Telescope, which is supported by the PPARC.

\end{document}